\documentclass[prb,aps,twocolumn,preprintnumbers,10pt,floats,amsmath,showpacs,amssymb,superscriptaddress]{revtex4-1}
\usepackage{epsfig}
\usepackage{amsmath}
\usepackage{amssymb}
\usepackage{amsfonts}
\usepackage{mathptmx}
\usepackage{dcolumn}
\usepackage{eucal}
\usepackage{bm}
\usepackage{color}
\usepackage[colorlinks,linkcolor=blue,citecolor=blue]{hyperref}

\usepackage{epstopdf}
\usepackage{graphicx}

\def \be{\begin{equation}}
\def \ee{\end{equation}}

\begin{document}

\title{Parasitic effects in SQUID-based radiation comb generators}
\author{R. Bosisio}
\email{riccardo.bosisio@nano.cnr.it}
\affiliation{SPIN-CNR, Via Dodecaneso 33, 16146 Genova, Italy}
\affiliation{NEST, Instituto Nanoscienze-CNR and Scuola Normale Superiore, I-56127 Pisa, Italy}
\author{F. Giazotto}
\email{giazotto@sns.it}
\affiliation{NEST, Instituto Nanoscienze-CNR and Scuola Normale Superiore, I-56127 Pisa, Italy}
\author{P. Solinas}
\email{paolo.solinas@spin.cnr.it}
\affiliation{SPIN-CNR, Via Dodecaneso 33, 16146 Genova, Italy}

\begin{abstract}
We study several parasitic effects on the implementation of a Josephson radiation comb generator (JRCG) based on a dc superconducting quantum interference device (SQUID) driven by an external magnetic field.
This system can be used as a radiation generator similarly to what is done in optics and metrology, and allows one to generate up to several hundreds of harmonics of the driving frequency.
First we take into account how assuming a finite loop geometrical inductance and junction capacitance in each SQUID may alter the operation of this device.
Then, we estimate the effect of imperfections in the fabrication of an array of SQUIDs, which is an unavoidable source of errors in practical situations.
We show that the role of the junction capacitance is in general negligible, whereas the geometrical inductance has a beneficial effect on the performance of the device. The errors on the areas and junction resistance asymmetries may deteriorate the performance, but their effect can be limited up to a large extent with a suitable choice of fabrication parameters.
\end{abstract}
\pacs{
74.50.+r, 
85.25.Dq, 
06.20.fb, 
04.40.Nr 
}
\maketitle
\section{Introduction}
Over the last decade, important advancements in the field of optical frequency combs have been reported~\cite{udem2002optical,cundiff2003,delhaye2007}. These have led to remarkable progresses in extending the accuracy of atomic clocks to the optical frequency region, with profound implications in several research areas, spanning from optical metrology~\cite{hansch1999laser} and high precision spectroscopy~\cite{bloembergen1977nonlinear,hansch1994frontiers} to telecommunication technologies~\cite{udem2002optical,foreman2007remote}.

In two recent papers~\cite{Solinas2014,Solinas2015} the implementation of radiation comb generators using dc superconducting quantum interference devices (SQUIDs) or extended Josephson junctions were discussed.
Assuming realistic experimental parameters, it was shown that such devices would be able to generate hundreds of harmonics of the driving frequency.
For example, at $200$ GHz a substantial output power of the order of a fraction of nW could be delivered using a standard $1$ GHz frequency drive.
This extraordinary frequency up-conversion opens the way to many applications from low-temperature microwave electronics to on-chip sub-millimeter wave generation.
The devices discussed in Refs.~\cite{Solinas2014,Solinas2015} were ``ideal'' in the sense that parasitic effects which can be present in a real structure were neglected. In light of a realistic implementation, such effects are unavoidable and must be taken into account.
\begin{figure}[t]
\centering
\includegraphics[width=0.95\columnwidth,  keepaspectratio]{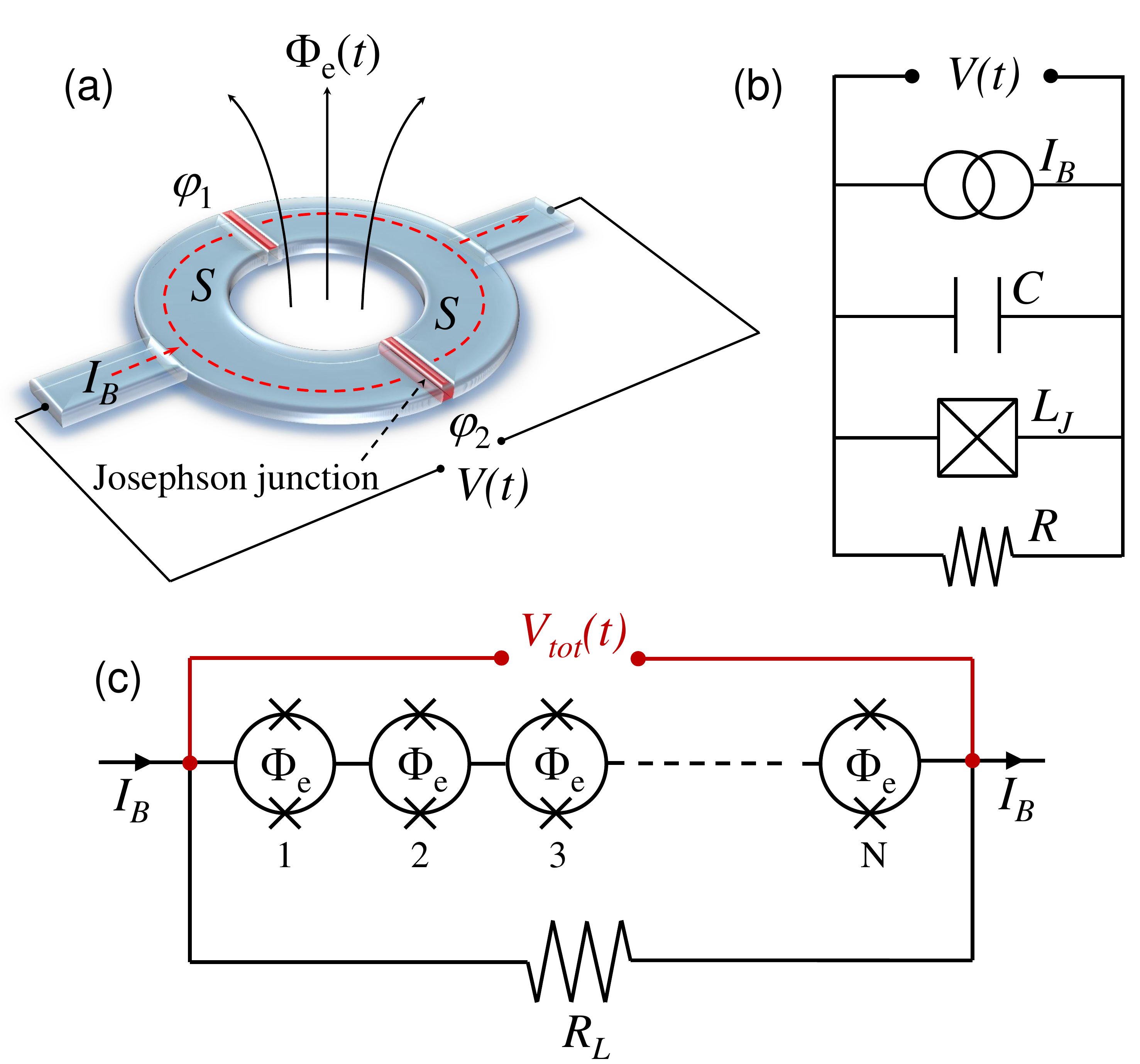}
\caption{(Color online) (a) Sketch of the single Josephson radiation comb generator, a SQUID subject to a time-dependent magnetic flux $\Phi_e (t)$ which induces voltage pulses $V(t)$ across the interferometer. The red regions denote  the  two  Josephson  tunnel  junctions, $I_B$ is the constant bias current, $\varphi_i$ is the superconducting phase across the $i$-th junction and $S$ are the superconducting electrodes. (b) RCSJ model circuit where $R$, $L_J$, and $C$ are the resistance, the Josephson inductance and the capacitance of the SQUID, respectively. (c) Sketch of a linear array of $N$ SQUIDs, connected together via a superconducting wire, and coupled to a load resistance $R_L$. Each SQUID is pierced by a uniform magnetic flux $\Phi$. The total voltage $V_\text{tot}(t)$ which develops across the array is given by the sum of all the voltage drops across each single SQUID.} 
\label{fig:SQUID_figuracompleta}
\end{figure}
In the present work we investigate extensively the impact of several parasitic effects on the phenomenology and performance of the SQUID-based radiation comb generator theoretically proposed in Ref.~\onlinecite{Solinas2014}. 
Namely, we analyze the case in which the SQUIDs have a finite loop geometrical inductance and junction capacitance, and then we estimate the role of adding uncertainty in the SQUIDs areas and asymmetry parameters when building up a chain.
We treat each one of these effects separately in order to emphasize their impact both on the physics and on the performance of our device.
In particular we show that the junction capacitance plays a negligible role for our choice of parameters, whereas the loop geometrical inductance has a beneficial effect on the performance of the device. On the other hand, the errors on the SQUID areas and junction resistance asymmetries may deteriorate the radiation comb generator performance, but their effect remains quite moderate if such errors are within a tolerance of $1\%$ and $0.5\%$ for the areas and the junction resistance asymmetry parameters, respectively.

The paper is structured as follows: First, we review the device theoretical analysis in Sec.~\ref{sec:model}. In Sec.~\ref{sec:results} we discuss how each parasitic effect alter the device performance: The role of a finite SQUID geometrical inductance and junction capacitance are investigated in Secs.~\ref{sec:results_inductance} and \ref{sec:results_capacitance}, respectively. Then, in Sec.~\ref{sec:results_areas} we estimate the impact of an uncertainty in the SQUIDs areas when adding them in series to build a linear array, whereas in Sec.~\ref{sec:results_asymmetries} we consider SQUIDs with different asymmetry parameters. A discussion about the experimental feasibility of the proposed system as well as the estimate of its realistic performance when all the aforementioned effects are taken into account at once are the content of Sec.~\ref{sec:feasibility}.
Finally, our conclusions are gathered in Sec.~\ref{sec:conclusions}.

\section{Theoretical analysis}
\label{sec:model}
In this section we briefly review the physical arguments leading to the prediction of the $\pi$-jumps of the superconducting phase, and the consequent generation of voltage pulses, using SQUID devices. Since these were extensively discussed in Refs.~\onlinecite{Solinas2014,Solinas2015} both for devices based on SQUIDs and on extended Josephson junctions, we recall here only the basic principles, without focusing on the details.
\subsection{Single SQUID}
We consider a SQUID biased by a constant current $I_B$ and driven by an external, time-dependent magnetic flux $\Phi_e(t)$ [see Fig.~\ref{fig:SQUID_figuracompleta}(a)].
Due to the first Josephson relation~\cite{Tinkham2012introduction}, the Josephson current through the SQUID is 
\be
I_J = I_{c1} \sin\varphi_1 +I_{c2} \sin\varphi_2, 
\ee
where $\varphi_i$ and $I_{ci}$ ($i$=1,2) are the phase across and the critical current of the $i$-th junction, respectively.
In the limit of negligible inductance\footnote{The effect of the loop inductance of the SQUID will be addressed in Sec.~\ref{sec:model}}, by introducing the superconducting phase across the SQUID $\varphi=(\varphi_1+\varphi_2)/2$ and using the flux quantization relation~\cite{Barone1982,Tinkham2012introduction} $\varphi_2-\varphi_1=-2\pi\Phi_e(t)/\Phi_0$, the current ($I_J$) vs phase relation of the SQUID can be written as
\be
I_J(\varphi;\phi)=I_+\,[\cos\phi\sin\varphi+r\,\sin\phi\cos\varphi],
\label{eq:IJ_SQUID}
\ee
where $\phi=\pi\Phi_e/\Phi_0$ ($\Phi_0=h/2e\simeq 2\times 10^{-15}$ Wb is the flux quantum), $I_+=I_{c1}+I_{c2}$, and $r=(I_{c1}-I_{c2})/(I_{c1}+I_{c2})$ expresses the degree of asymmetry of the interferometer. 
Equation~\eqref{eq:IJ_SQUID} describes the well-known oscillations of the SQUID critical current $I_c(\phi)=\text{max}_\varphi I_J(\varphi;\phi)$ as a function of the magnetic flux~\cite{Tinkham2012introduction}, with minima occurring at integer multiples of $\Phi_0/2$. 

For a fixed bias current, when $\Phi_e$ crosses a critical-current minimum we see from Eq.~\eqref{eq:IJ_SQUID} that a change of sign in $\cos
\phi$ must be accompanied by a change of sign in $\sin\varphi$ in order for the current to maintain its direction. This is accomplished by a phase jump of $\pi$~\cite{Solinas2014,Solinas2015,giazotto2013coherent,martinez2012nature,martinez2014quantum} which, owing to the second Josephson relation~\cite{Tinkham2012introduction}, results in a voltage pulse $V(t)$ across the SQUID.
The physical origin of the $\pi$-jump of the superconducting phase can be also easily understood on an energetic ground.
For a symmetric ($r=0$) SQUID in the absence of any bias current, the time-dependent Josephson potential is $E_J(t)=\int I_J V(t)dt = -E_{J0}f(t)\cos\varphi$, where $f(t)=\cos(\phi)$ and $E_{J0}=\Phi_0^2\nu\alpha/(2\pi R)$.
At the initial time ($t=0$) this potential has minima at $\varphi=2k\pi$ (with $k$ integer).
When the magnetic flux reaches the diffraction node at $\Phi_e=\Phi_0/2$, $E_J$ vanishes, and for $\Phi_e>\Phi_0/2$ $f(t)$ changes its sign.
The former equilibrium points $\varphi=2k\pi$ have become unstable and hence, to remain in a minimum energy state, $\cos\varphi$ must change sign, meaning that the superconducting phase must undergo a $\pi$-jump to reach a new minimum at $\varphi=(2k+1)\pi$.
Notice that a finite bias current $I_B$ is then necessary to induce a preferential direction to the phase jumps.

To determine the details of the voltage pulses, we rely on the so-called resistively and capacitively shunted Josephson junction (RCSJ) model~\cite{Tinkham2012introduction, gross2005applied} adapted to a SQUID [see Fig.~\ref{fig:SQUID_figuracompleta}(b)], in which each Josephson junction is modelized as a circuit with a capacitor $C$, a resistor $R$, and a non-linear (Josephson) inductance $L_J$ arranged in a parallel configuration. 
We consider an external sinusoidally-driven magnetic flux with frequency $\nu$ and amplitude $\epsilon$, centered in the first node of the interference pattern, so that 
\be
\Phi_e(t) = \frac{\Phi_0}{2}[1-\epsilon  \cos (2 \pi \nu t )].
\ee
As a result, the magnetic flux crosses the nodes of the interference pattern at $t_k= (2k+1)/ 4 \nu$, with $k$ integer.
The equation of motion for $\varphi$ can be written as:
\begin{equation}
   \frac{\hbar C}{2e} \ddot{\varphi} + \frac{\hbar}{2eR} \dot{\varphi} +I_+ f(\varphi,t) = I_B,
  \label{eq:RCSJ_dim}
\end{equation}
where $C$ is the junction capacitance, $R$ is the total shunting resistance of the SQUID, $I_B$ is the external bias current and $f(\varphi,t)=I_J[\varphi;\phi(t)]/I_+$.
This equation can be expressed in terms of the dimensionless time $\tau = 2 \pi \nu t$. Recalling $\hbar/(2 e)= \Phi_0/ 2 \pi$, we obtain~\cite{Solinas2014}
\begin{equation}
   c \frac{d^2 \varphi}{d \tau^2}+  \frac{d \varphi}{d \tau} - \alpha[ f(\varphi,\tau) - \delta]=0,
  \label{eq:RCSJ_adim}
\end{equation}
where $c =2 \pi R C \nu$, $\alpha= I_+ R/(\Phi_0 \nu)$, and $\delta=I_B/I_+\ll 1$ is the dimensionless bias current.

The ability to generate a sequence of voltage pulses suggests an application similar to the frequency combs used in optics~\cite{udem2002optical,delhaye2007}. In this context, the most relevant feature becomes the sharpness of the voltage pulse, which is related to the number of harmonics generated. The sharpness is essentially determined by the product $I_+R$, which in turn depends on the material properties of the Josephson junctions~\cite{Solinas2014}.
\subsection{SQUID array with load resistor}
So far the analysis has been focused on the voltage produced by a single JRCG in the absence of any external load.
A quantity of experimental relevance is the \emph{extrinsic} power that can be transferred to a load resistance $R_L$.
Although the total output power provided by a single SQUID is fairly small, it can be boosted by using a linear array of $N$ SQUIDs, connected together via a superconducting wire [see Fig.~\ref{fig:SQUID_figuracompleta}(c)]. A similar approach is used for the realization of the metrological standard for voltage based on the Josephson effect~\cite{udem2002optical,delhaye2007,hansch1999laser}.
If we neglect the coupling among the SQUIDs via mutual inductance and/or cross capacitance and inductance of the superconducting wire (see Sec.~\ref{sec:feasibility}), the dynamics of each SQUID is independent from the rest of the array~\cite{Solinas2014}.
In this case the total voltage produced by the chain is given by summing up the voltages developed across each single SQUIDs:
\be
V_{\text{tot}}(t)=\sum_{i=1}^N\,V_i(t).
\ee
Assuming for simplicity that the $N$ SQUIDs are identical, this can be rewritten as $V_{\text{tot}}(t)=N V(t)$: This is the voltage which develops across the load resistor $R_L$ [see Fig.~\ref{fig:SQUID_figuracompleta}(c)].
As a consequence, a current $I_L=N V(t)/R_L$ flows across it, having denoted $R_L$ the (real) impedance of the load.
The bias current $I_B$ is thus split into two parts, one entering the SQUID array, the other ($I_L$) flowing through the load resistor. 
This is accounted for by replacing the resistance $R$ ``seen'' by each SQUID in Eq.~\eqref{eq:RCSJ_dim} with an effective resistance
\begin{equation}
   R_\text{eff}=\left(\frac{1}{R}+\frac{N}{R_L}\right)^{-1}=\frac{RR_L}{R_L+NR},
  \label{eq:R_eff}
\end{equation}
and Eq.~\eqref{eq:RCSJ_dim3} becomes then
\begin{equation}
   \frac{\hbar C}{2e} \ddot{\varphi} + \frac{\hbar}{2eR_\text{eff}} \dot{\varphi} +I_+ f(\varphi,t) = I_B.
  \label{eq:RCSJ_dim3}
\end{equation}
This effective change in the shunt resistance modifies the dynamics of each single SQUID.
In particular, being $R_\text{eff}<R$, it also reduces the power $P=N^2V^2/R_L$ that can be delivered to the load.
Since for a single SQUID~\cite{Solinas2014} $V^2\propto R_\text{eff}$, using Eq.~\eqref{eq:R_eff} we find that $P\propto N^2$ for $N\ll R_L/R$, whereas $P\propto N$ for $N\gg R_L/R$.

We will now turn to the analysis of different parasitic effects on the performance of this system. Each one of these effects will be first treated independently, in order to better emphasize its impact on the physics of the device. After that, we will try to give a more realistic estimate of the performance by considering all these effects at once. All the results that follow are for an array of $N=50$ SQUIDs made of Nb/AlOx/Nb Josephson junctions~\cite{patel1999self}, under a 1GHz magnetic flux driving frequency. In particular, we have set the junction shunt resistance $R=20$ Ohm, the effective resistance $R_\text{eff}\simeq 1$ Ohm (estimated using Eq.~\eqref{eq:R_eff} for a linear array of $N=50$ SQUIDs with a load resistance $R_L$=50 Ohm), the SQUID critical current $I_+=100~\mu$A, the bias current $I_B=10^{-3}I_+$ and the amplitude of the magnetic flux oscillations $\varepsilon$ = 0.9. Moreover, the typical size of the Nb/AlOx/Nb Josephson junctions is at most $1\,\mu\rm{m}\times 1\,\mu\rm{m}$, and the size of each SQUID is of the order of $1\,\mu\rm{m}-10\,\mu\rm{m}$.




\section{Results}
\label{sec:results}

\subsection{Finite SQUID geometrical inductance}
\label{sec:results_inductance}
In this section we investigate how taking into account a finite loop geometrical inductance of an \emph{individual} SQUID modifies its dynamics under the effect of a time-dependent magnetic field.
Labeling ``1'' and ``2'' the two SQUID arms, and denoting $i_1(t)$ and $i_2(t)$ the (time-dependent) currents through each of them, we define the \emph{total} current $I$ trough the SQUID and the \emph{circulating} supercurrent $i_S$ as~\cite{Barone1982,Tinkham2012introduction}:
\begin{align}
I &= i_1(t)\,+\,i_2(t),  \cr
i_S &= [i_2(t)\,-\,i_1(t)]\,/\,2.
\label{eq:I_iS_1}
\end{align}
For a symmetric SQUID ($r=0$), the currents $i_1(t)$ and $i_2(t)$ are related to the voltage drops across the junctions and to the Josephson supercurrents by
\begin{align}
i_1(t) & = I_0\sin\varphi_1(t)+\frac{V_1}{R},  \cr
i_2(t) & = I_0\sin\varphi_2(t)+\frac{V_2}{R},
\label{eq:i1i2} 
\end{align}
where the time dependence of the superconducting phases is given by the second Josephson relation $d\varphi_k/dt=(2e/\hbar)\,V_k$ ($k=1,2$), $V_k$ being the electric potential difference across the $k$-th junction. 
Following De Waele~\cite{DeWaele1969} we neglect the dissipative contribution to the circulating supercurrent, proportional to $(V_2-V_1)/R$. This is a reasonable assumption up to driving frequencies $\nu$ of the order of the GHz~\cite{DeWaele1969}, and means that there is no appreciable contribution to $i_S$ originating from the Lenz's law of induction. Under this approximation Eqs.~\eqref{eq:I_iS_1} reduce to
\begin{align}
&I = 2I_0\,\sin\left(\frac{\varphi_1+\varphi_2}{2}\right) \cos\left(\frac{\varphi_1-\varphi_2}{2}\right)\,+\,2\frac{V(t)}{R}, \cr
&i_S = I_0\,\cos\left(\frac{\varphi_1+\varphi_2}{2}\right) \sin\left(\frac{\varphi_2-\varphi_1}{2}\right),
\label{eq:I_iS_2}
\end{align}
where $V(t)=[V_1(t)+V_2(t)]/2$ is the voltage drop generated across the SQUID.
In writing the flux quantization~\cite{Barone1982,Tinkham2012introduction}:
\be
\varphi_2-\varphi_1=-2\pi\frac{\Phi(t)}{\Phi_0},
\label{eq:flux_quantization}
\ee
now the \emph{total} magnetic flux piercing the SQUID is $\Phi(t)=\Phi_e(t)+L_g i_S$, which differs from the external (time-dependent) term $\Phi_e(t)$ because of the geometrical inductance of the loop $L_g$.
Using Eqs.~\eqref{eq:I_iS_2} and~\eqref{eq:flux_quantization}, after some straightforward algebra, we can express the total current through the SQUID and the total magnetic flux as:
\begin{subequations}
\begin{align}
&I = 2I_0\,\sin\varphi(t) \cos\left(\frac{\pi\Phi(t)}{\Phi_0}\right)\,+\,2\frac{V(t)}{R},\label{eq:I_iS_3a} \\
&\Phi(t) = \Phi_e(t) \,-\, L_g I_0 \sin\left(\frac{\pi\Phi(t)}{\Phi_0}\right)\cos\varphi(t),\label{eq:I_iS_3b}
\end{align}
\end{subequations}
where the phase $\varphi(t)= [\varphi_1(t)+\varphi_2(t)]/2$ is related to the voltage drop across the SQUID via $V(t)=(\hbar/2e)\,d\varphi(t)/dt$.
Equation~\eqref{eq:I_iS_3b} offers the following physical interpretation: At any instant of time $t$, the finite loop inductance modifies the external flux $\Phi_e(t)$ piercing the SQUID, and the resulting total magnetic flux $\Phi(t)$ has to be evaluated self-consistently. Once this is done, the dynamics of the SQUID phase $\varphi(t)$ [as well as the total voltage drop across the device $V(t)$] can be evaluated via Eq.~\eqref{eq:I_iS_3a}.

In writing $V(t)$ as the mean voltage generated across the two junctions, we have implicitly assumed that the two SQUID arms have the same inductance. Accounting for different arms inductances $L_1\neq L_2$ would result in an additional correction to the magnetic flux piercing the SQUID, which would become\cite{Barone1982,Tinkham2012introduction} $\Phi = \Phi_e + L_g i_S - \alpha_L L_gI/2$, where $L_g=L_1+L_2$, $\alpha_L = (L_1-L_2)/L_g$, while $I$ and $i_S$ are defined by Eqs.~\eqref{eq:I_iS_2}.
From this expression we see that unless the difference between $L_1$ and $L_2$ is large (i.e., comparable to $L_g$) the term $\propto \alpha_L L_gI/2$ is a minor correction to the magnetic flux, with respect to $L_g i_S$.

In order to quantify the effect of the inductance, we have solved numerically the RCSJ-equation~\eqref{eq:RCSJ_dim3} for an array of 50 symmetric ($r=0$) SQUIDs made of Nb/AlOx/Nb junctions~\cite{patel1999self}, and computed the voltage pulses for different values of the loop geometrical inductance, compatible with typical SQUID dimensions. 

\begin{figure}[b]
\centering
\includegraphics[width=0.9\columnwidth,  keepaspectratio]{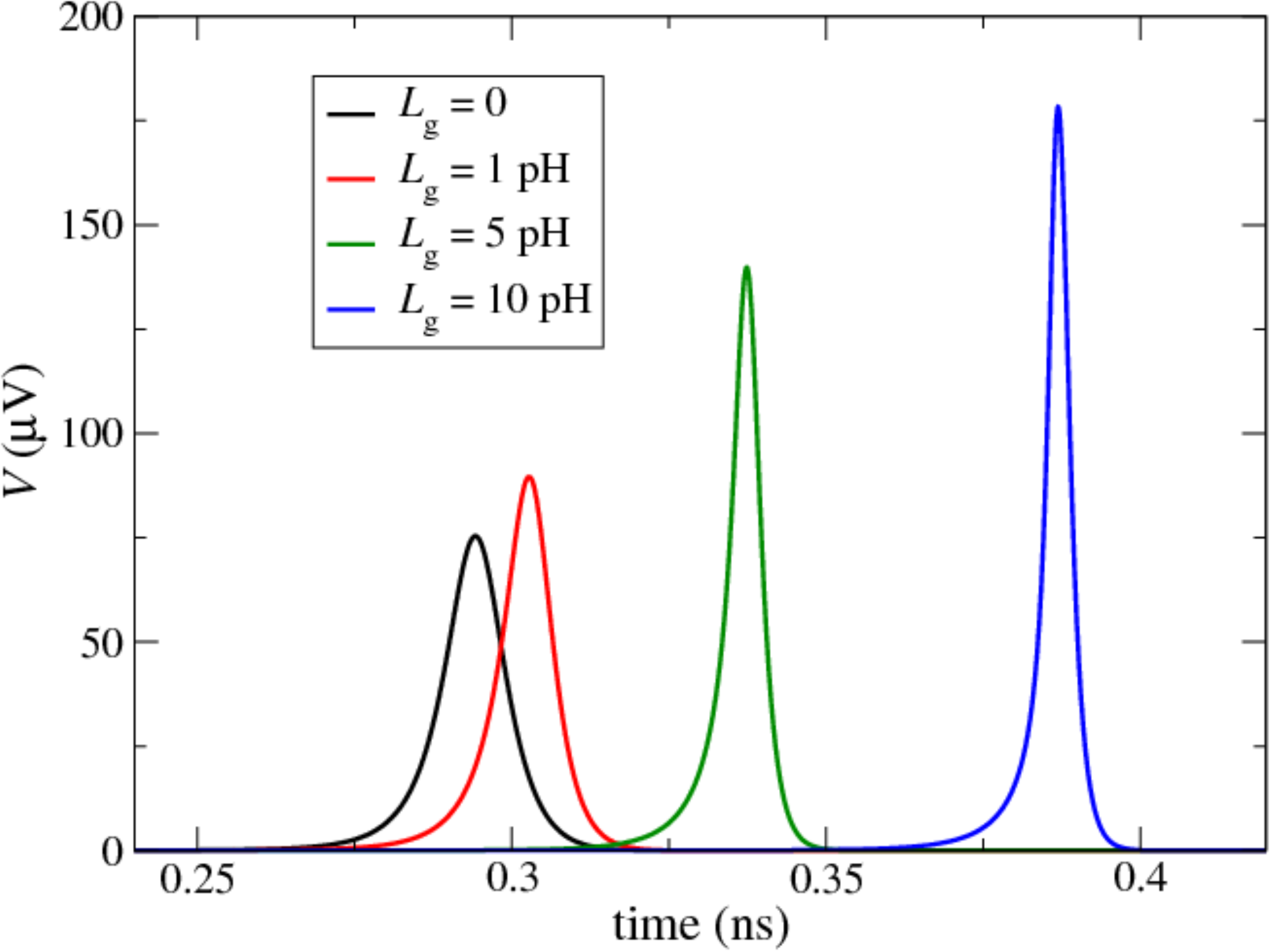}
\caption{(Color online) Behavior of a typical voltage pulse generated by each symmetric Nb/AlOx/Nb SQUID ($r=0$) of the array, for different values of its (geometrical) inductance $L_g$. The driving frequency is $\nu$=1 GHz, whereas the other parameters are those typical of a Nb/AlOx/Nb Josephson junction~\cite{patel1999self}, given at the end of Sec.~\ref{sec:model}. 
} 
\label{fig:SQUID_Nb_inductance_effect_nu1}
\end{figure}

\begin{figure}[h]
\centering
\includegraphics[width=0.9\columnwidth,  keepaspectratio]{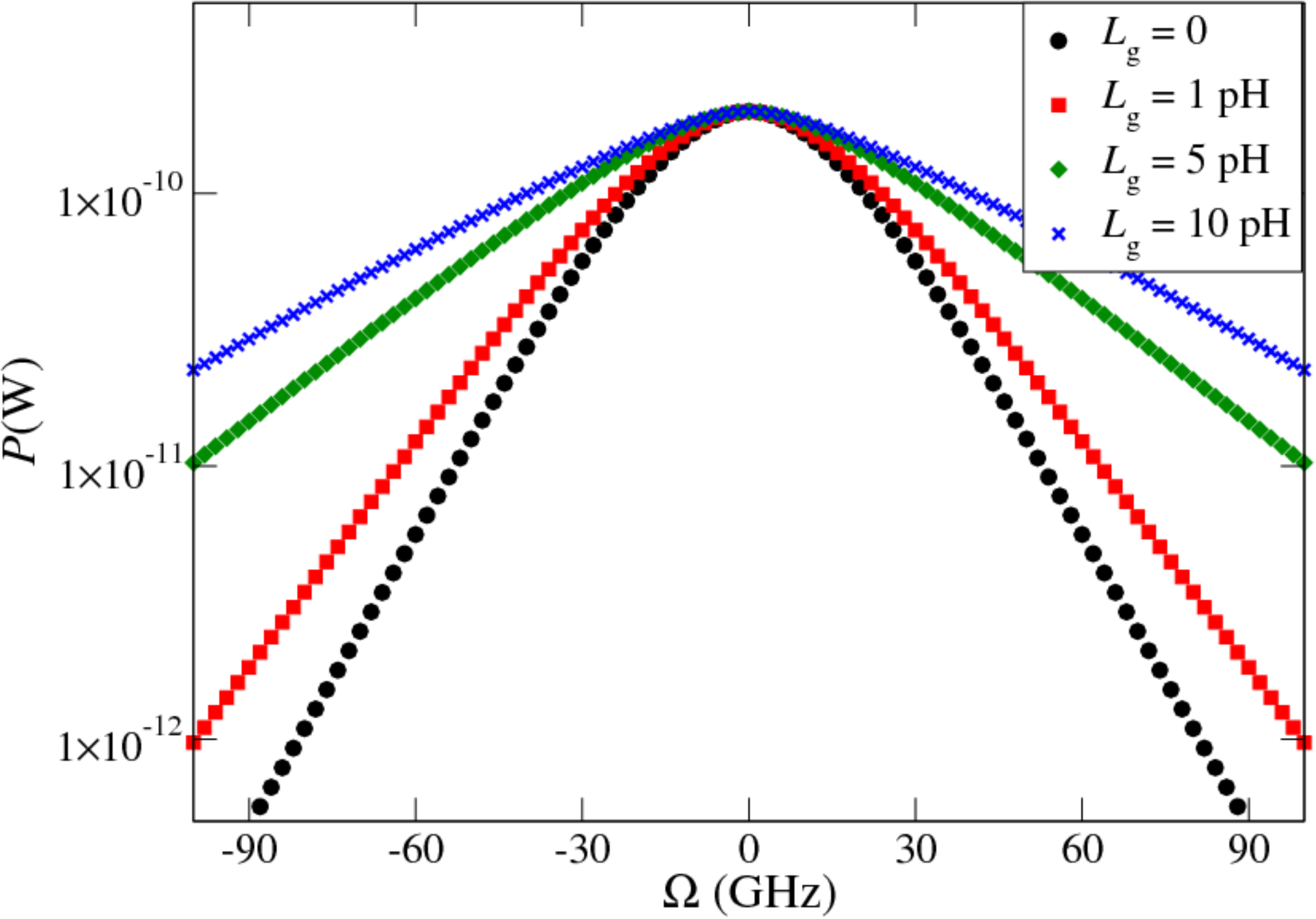}
\caption{(Color online) Power spectrum of the SQUID-based radiation comb generator over a 50 Ohm transmission line. Different symbols correspond to different values of the geometrical inductance $L_g$. The calculation is performed for a $N = 50$ chain of nominally identical and symmetric Nb/AlOx/Nb SQUIDs, subject to a $\nu$ = 1 GHz driving. The parameters are the same as in Fig.~\ref{fig:SQUID_Nb_inductance_effect_nu1}. Notice that only the even harmonics of the driving frequency are shown, the contribution associated to the odd ones being vanishingly small.} 
\label{fig:SQUID_Nb_inductance_power_nu1}
\end{figure}

In Fig.~\ref{fig:SQUID_Nb_inductance_effect_nu1} we show the effect of a finite inductance on the shape of the voltage pulse generated by each SQUID of the chain.
We notice that a geometrical inductance $L_g$ of the order $\sim$pH is a reasonably good assumption for a SQUID with radius $r$ of the order $\sim \mu$m if we approximate $L_g\simeq\mu_0 r$, $\mu_0$ being the vacuum permeability.
From the figure we see that the principal effect of increasing $L_g$ is that the voltage pulses are delayed with respect to the $L_g=0$ case, and furthermore they are sharper and higher. This is a direct consequence of the change in the time-dependent magnetic flux profile. Indeed, starting at $t=0$, it turns out that $\Phi(t)$ is initially reduced by virtue of the second term in Eq.~\eqref{eq:I_iS_3b}.
This means that the condition at which the $\pi$-jump of the phase is met (that is, $\Phi(t)=\Phi_0/2$) is verified at a later time than $t_k$ (see Sec.~\ref{sec:model}), and the same holds for the voltage pulse.
In addition, the fact that the shape of $\Phi(t)$ is altered from the original cosinusoidal profile induces a faster relaxation of the phase $\varphi$ toward the energy minimum.
As a consequence, the voltage peaks for finite geometrical inductance are sharper and skewed with respect to the $L_g=0$ case (leftmost curve in Fig.~\ref{fig:SQUID_Nb_inductance_effect_nu1}).
This has a beneficial impact on the emitted radiation spectrum $P(\Omega)$, as it is confirmed in Fig.~\ref{fig:SQUID_Nb_inductance_power_nu1}, where we show the power generated by a chain of $N=50$ nominally identical and symmetric SQUIDs made of Nb/AlOx/Nb junctions, driven by a 1 GHz oscillating magnetic field, for different values of the loop geometrical inductance $L_g$.
As we can see, the device with $L_g$=10 pH is able to provide a power of about 0.1 nW at 20 GHz (corresponding to the 20-th harmonics of the driving frequency). Notice finally that only the even harmonics of the driving frequency are shown in the power spectrum of the emitted radiation, the contribution of the odd ones being vanishingly small for symmetric ($r=0$) SQUIDs~\cite{Solinas2014}.
%
%
\subsection{Finite SQUID junction capacitance}
\label{sec:results_capacitance}
In this section we investigate the effect of taking into account a finite SQUIDs junction capacitance. 
In order to do this, we have solved the differential RCSJ equation [Eq.~\eqref{eq:RCSJ_dim3}] for the SQUID phase dynamics without neglecting the second-order (diffusive) term. Details on the numerical procedure are given in appendix~\ref{sec:app_capacitance}.

In Fig.~\ref{fig:SQUID_Nb_capacitance_effect_nu1} we show how the typical voltage pulse generated by each SQUID of the chain is altered due to the effect of a finite junction capacitance $C$. 
We notice that increasing $C$ up to $1$ pF has the only effect of making the voltage peak slightly skewed and sharper: This would be beneficial in terms of output power.
For larger values of the junctions capacitance, the second order term in Eq.~\eqref{eq:RCSJ_dim3} becomes more important and the system starts operating in the under-damped regime. This is evident for $C=2.5$ pF (rightmost curve in Fig.~\ref{fig:SQUID_Nb_capacitance_effect_nu1}), at which the voltage $V(t)$ exhibits small oscillations before relaxing to zero, taking also negative values.
However, all these effects would be relevant for large Josephson junctions, whereas in this work we focus rather on small Nb/AlOx/Nb junctions, typically characterized by a relatively low capacitance ($C\lesssim 100$ fF). In this case, we see from Fig.~\ref{fig:SQUID_Nb_capacitance_effect_nu1} that there is no appreciable difference with respect to the zero-capacitance case (the corresponding curves are essentially indistinguishable). As a consequence, our device operates always in the over-damped regime~\cite{Tinkham2012introduction, gross2005applied}.
According to these results, we do not expect any relevant modifications in the power spectrum of the emitted radiation with respect to the ideal (zero capacitance) case, and thus we decided not to show it. 

In addition, we have also performed numerical simulations taking into account the combined effect of both a finite junction capacitance \textit{and} loop inductance, but we did not observe any relevant modification with respect to the results discussed in this and the previous subsection~\ref{sec:results_inductance}. 
\begin{figure}[t]
\centering
\includegraphics[width=0.9\columnwidth,  keepaspectratio]{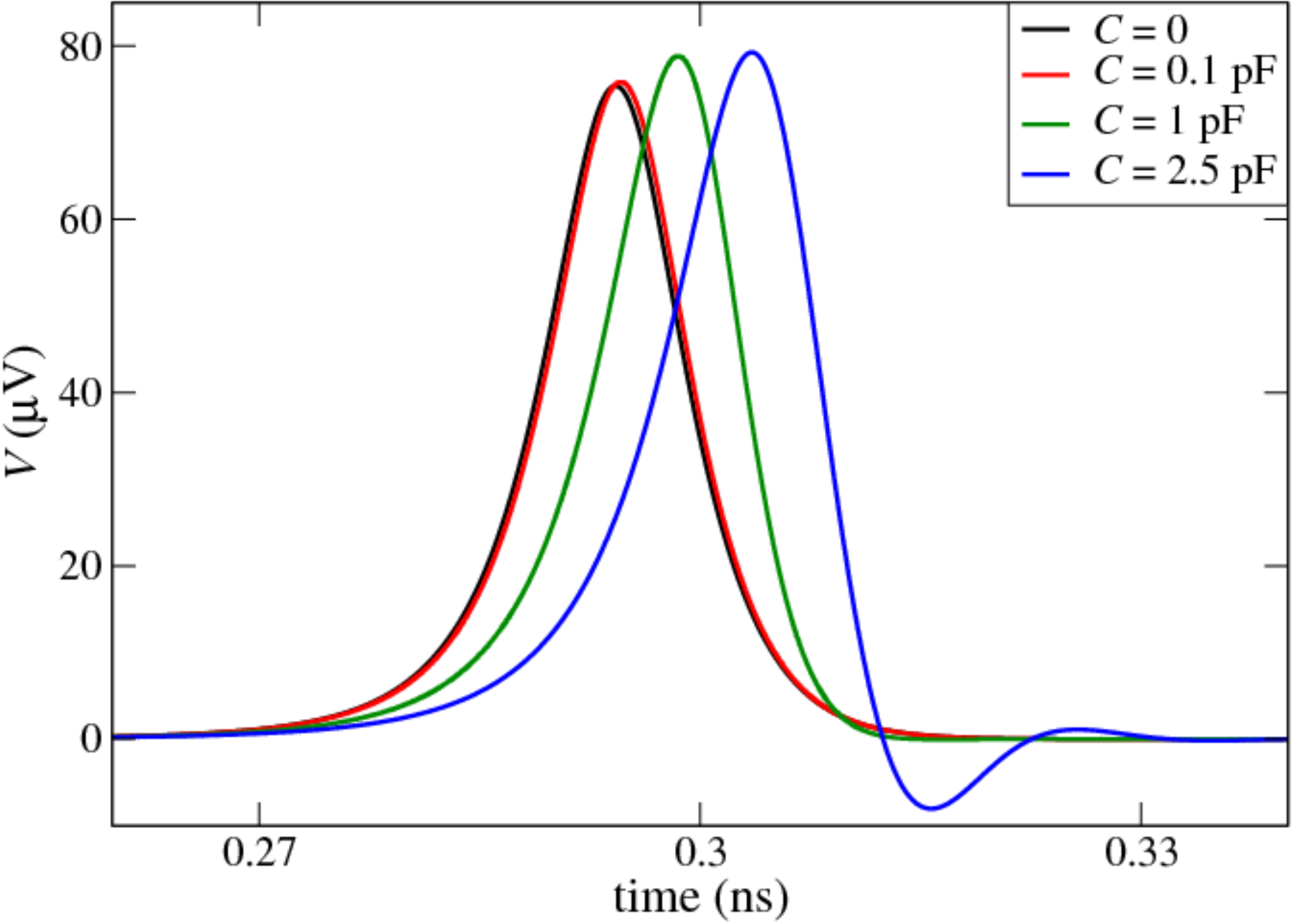}
\caption{(Color online) Behavior of a typical voltage pulse generated by each symmetric Nb/AlOx/Nb SQUID ($r=0$) of the array, for different values of the junction capacitance $C$. The driving frequency is $\nu$=1 GHz, whereas the SQUIDs parameters are the same as in the previous figures.} 
\label{fig:SQUID_Nb_capacitance_effect_nu1}
\end{figure}
%
%
%
%
%
%
\subsection{Uncertainty on the SQUIDs areas}
\label{sec:results_areas}
When fabricating an array of $N$ SQUIDs, it is most unlikely to be able to make them all identical. Inevitable imprecisions in the lithographic processes imply that the SQUIDs will have slightly different areas. As a consequence, if the array is embedded in a coil which generates an ideally uniform magnetic field, the resulting flux $\Phi_e$ piercing each SQUID of the array will be different: Larger SQUIDs will be pierced by a larger magnetic flux, and vice-versa.
This will induce a shift in the time at which the condition $\Phi_e=\Phi_0/2$ (when the superconducting phase experiences a $\pi$-jump) is met: The phase will jump earlier in larger SQUIDs.

To better quantify this effect, let us associate a gaussian statistical distribution for the SQUID areas:
\be
A=A_0\,(1+\zeta_A)\quad \text{with} \quad \mathcal{P}(\zeta_A)=\frac{1}{\sqrt{2\pi}\sigma_A}\,\text{exp}\left(-\frac{\zeta_A^2}{2\sigma_A^2}\right),
\label{eq:AreasDistribution}
\ee
where $\zeta_A$ is a dimensionless parameter quantifying the degree of uncertainty on the SQUIDs areas, being normally distributed around zero with variance $\sigma_A^2$, whereas $A_0$ is the reference value for the surface delimited by the SQUID loop.
The standard deviation $\sigma_A$ can thus be seen as the percentage error within which the value of the area is known.
We can write the external magnetic flux as:
\begin{align}
\Phi_{e}(t) & =B(t)\,A = (B_0-B_1\,\cos(2\pi\nu t))\,A_0(1+\zeta_A) =\cr
            & =B_0 A_0\left(1-\frac{B_1}{B_0}\cos(2\pi\nu t)\right)\,(1+\zeta_A) =\cr
						& =\frac{\Phi_0}{2}(1+\zeta_A)\,\left[1-\varepsilon \,\cos(2\pi\nu t)\right],
\end{align}
where we defined $B_0 A_0\equiv \Phi_0/2$ and $\varepsilon\equiv B_1/B_0$.
The phase jump occurs at $\Phi_e(t)=\Phi_0/2$, that is, at a \emph{switch time} $\bar{t}$ determined by:
\begin{align}
& \zeta_A -\varepsilon(1+\zeta_A)\cos(2\pi\nu \bar{t})=0, \cr
 \to \,\,\,\,& \bar{t}=\frac{1}{2\pi\nu}\arccos\left(\frac{1}{\varepsilon}\frac{\zeta_A}{1+\zeta_A}\right)+k\pi,
\end{align}
where $k$ is a non-negative integer.
For sufficiently small $\zeta_A$, the above expression for $\bar{t}$ simplifies to:
\be
\bar{t}\approx \frac{1}{4\nu}(1+2k)-\frac{\zeta_A}{2\pi\nu\varepsilon} \equiv t_k - \frac{\zeta_A}{2\pi\nu\varepsilon},
\label{eq:switchtime}
\ee
where, as in Sec.~\ref{sec:model}, we have defined $t_k=(1+2k)/4\nu$. From this expression it is evident that larger SQUIDs ($\zeta_A>0$) switch before ($\bar{t}<t_k$), and vice-versa.
Notice also that, since the relation between $\zeta_A$ and $\bar{t}$ is linear, we can understand this result in terms of the distribution of the switch-times $\mathcal{P}(\bar{t})$, which can be easily computed:
\be
\mathcal{P}(\bar{t})=\frac{1}{\sqrt{2\pi}\lambda_A}\,\text{exp}\left(-\frac{(\bar{t}-t_k) ^2}{2\lambda_A^2}\right),
\ee
with $\lambda_A=\sigma_A/2\pi\varepsilon\nu$. This can be interpreted by stating that the times $\bar{t}$ at which the phase of the SQUIDs undergo a $\pi$-jump is normally distributed around $t_k$ with a variance $\lambda_A$ which is directly proportional to the uncertainty $\sigma_A$ on the SQUIDs areas.

\begin{figure}[b]
\centering
\includegraphics[width=0.9\columnwidth,  keepaspectratio]{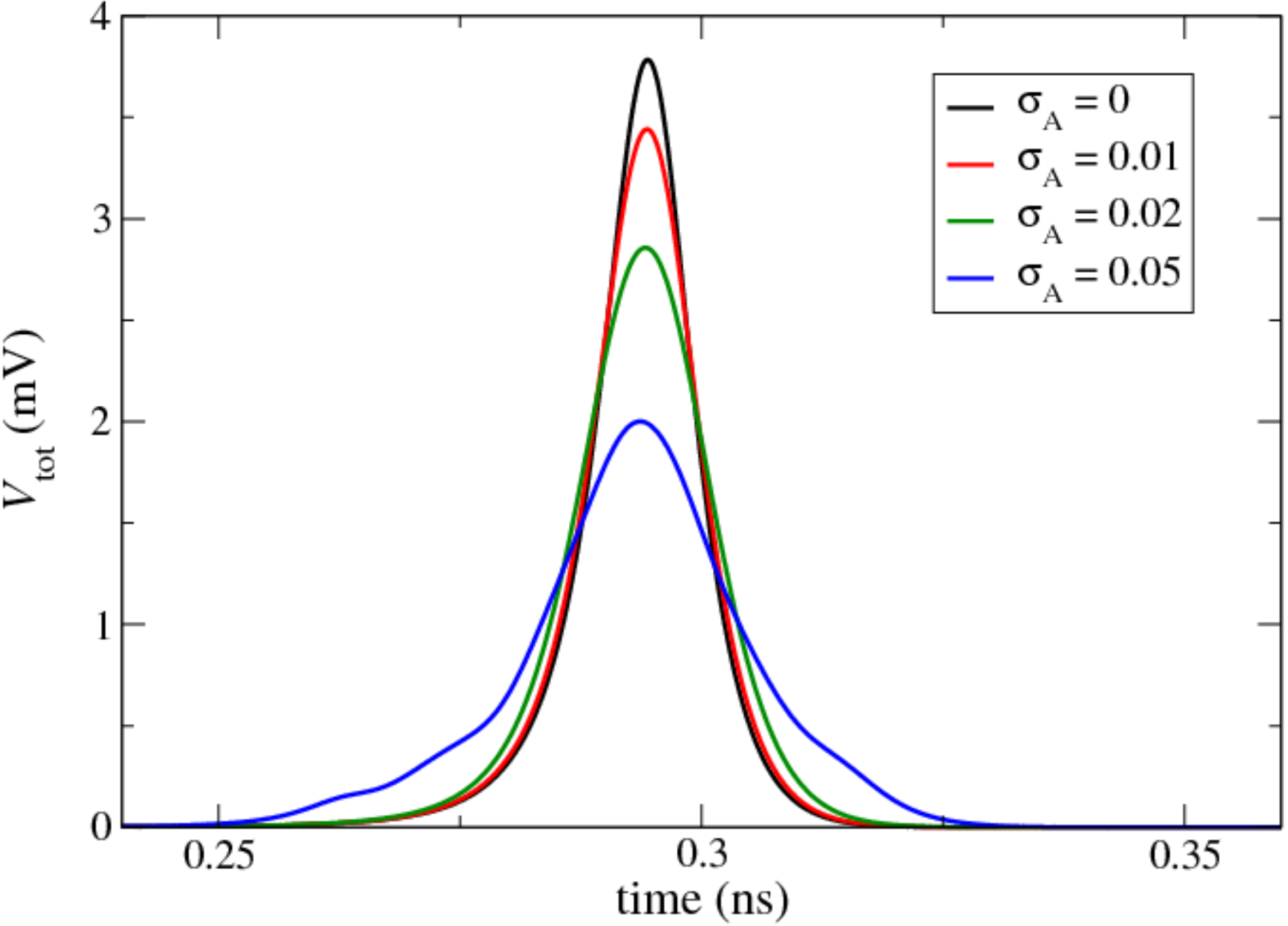}
\caption{(Color online) Behavior of a \emph{typical} sequence of two voltage pulses generated by an array of $N=50$ SQUIDs made of Nb/AlOx/Nb junctions with areas statistically distributed according to Eq.~\eqref{eq:AreasDistribution} for different values of the standard deviation $\sigma_A$. The driving frequency is $\nu$=1 GHz, whereas the SQUIDs parameters are the same as in the previous figures.} 
\label{fig:SQUID_Nb_errAreas_effect_nu1}
\end{figure}

\begin{figure}[h]
\centering
\includegraphics[width=0.9\columnwidth,  keepaspectratio]{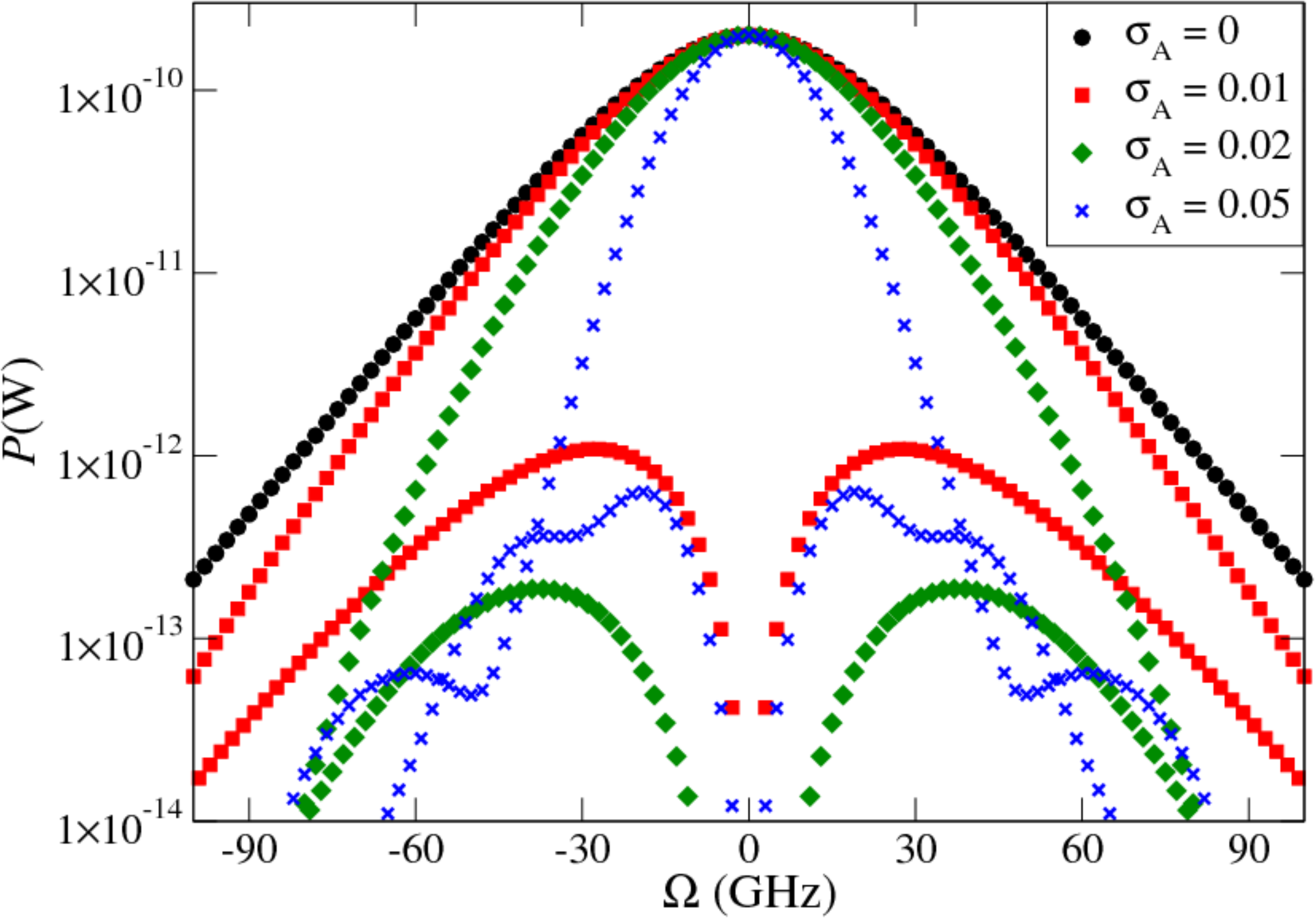}
\caption{(Color online) Average power spectrum of the SQUID-based radiation comb generator over a 50 Ohm transmission line, for different values of the standard deviation $\sigma_A$ of the areas distributions. The calculation is performed for a $N = 50$ chain of Nb/AlOx/Nb SQUIDs subject to a $\nu$=1 GHz driving.} 
\label{fig:SQUID_Nb_errAreas_power_nu1}
\end{figure}

In Fig.~\ref{fig:SQUID_Nb_errAreas_effect_nu1} we show how a \emph{typical} voltage pulse generated by a linear array of $N=50$ symmetric SQUIDs is altered by assuming different uncertainties $\sigma_A$ on the areas, up to five percent. By ``typical'' we mean that we have first computed $V_\text{tot}(t)$ for a single array of SQUIDs with random areas [according to Eq.~\eqref{eq:AreasDistribution}], and then we have iterated this procedure for many realizations of the array. 
We have finally calculated the average voltage pattern, and defined it as the typical one (see appendix~\ref{sec:app_statisticalapproach}).
We notice that the main effect is that the voltage peaks are broadened and lowered, due to the fact that a certain number of SQUIDs switch before and after $t_k$, the reference switch time for a SQUID of area $A_0$ [see Eq.~\eqref{eq:switchtime}]. 
As a consequence, the power spectrum of the emitted radiation is lowered,  exhibiting an exponential cut-off at high frequency.
Despite this, we notice in Fig.~\ref{fig:SQUID_Nb_errAreas_power_nu1} that this reduction is still very moderate for an uncertainty $\sigma_A=0.01$, in which case the power is reduced by less than one order of magnitude around $100$ GHz (corresponding to the $100$-th harmonics of the driving frequency), whereas it is basically unchanged at $20$ GHz. By increasing the error to $\sigma_A=0.05$, on the other hand, the power $P$ is reduced in a substantial way. 
We finally note that, in contrast to Fig.~\ref{fig:SQUID_Nb_inductance_power_nu1}, in the power spectra for $\sigma_A\geq 0.01$ the non-dominant (odd) harmonics are visible (bottom curves).
Remarkably, they show complex structure when increasing $\sigma_A$. This is evident for $\sigma_A=0.05$: In this case, for $\Omega\gtrsim 40$ GHz, the power associated to odd harmonics becomes of the same order, if not larger, than that associated to the odd ones.
%
%
%
%
\subsection{Uncertainty on the SQUIDs asymmetry parameters}
\label{sec:results_asymmetries}
Another possible source of non-ideality in the fabrication of an array of SQUIDs stems from the asymmetry between the two Josephson junctions composing each element of the array.
This is quantified in terms of the asymmetry parameter $r=(I_{c,1}-I_{c,2})/(I_{c,1}+I_{c,2})$, as explained in Sec.~\ref{sec:model}.
We notice that assuming a statistical symmetric distribution for the parameter $r$ around 0 (corresponding to an ideally symmetric SQUID) would be much detrimental for the device performance, because SQUIDs with $I_{c,2}>I_{c,1}$ generate opposite voltage pulses with respect to SQUIDs with $I_{c,2}<I_{c,1}$, for small bias current~\cite{Solinas2014}. Thus, when summing up all the pulses to compute the total voltage, the contributions associated to $r>0$ would basically compensate those associated to $r<0$, resulting in a poor performance in terms of output power.

\begin{figure}[b]
\centering
\includegraphics[width=0.9\columnwidth,  keepaspectratio]{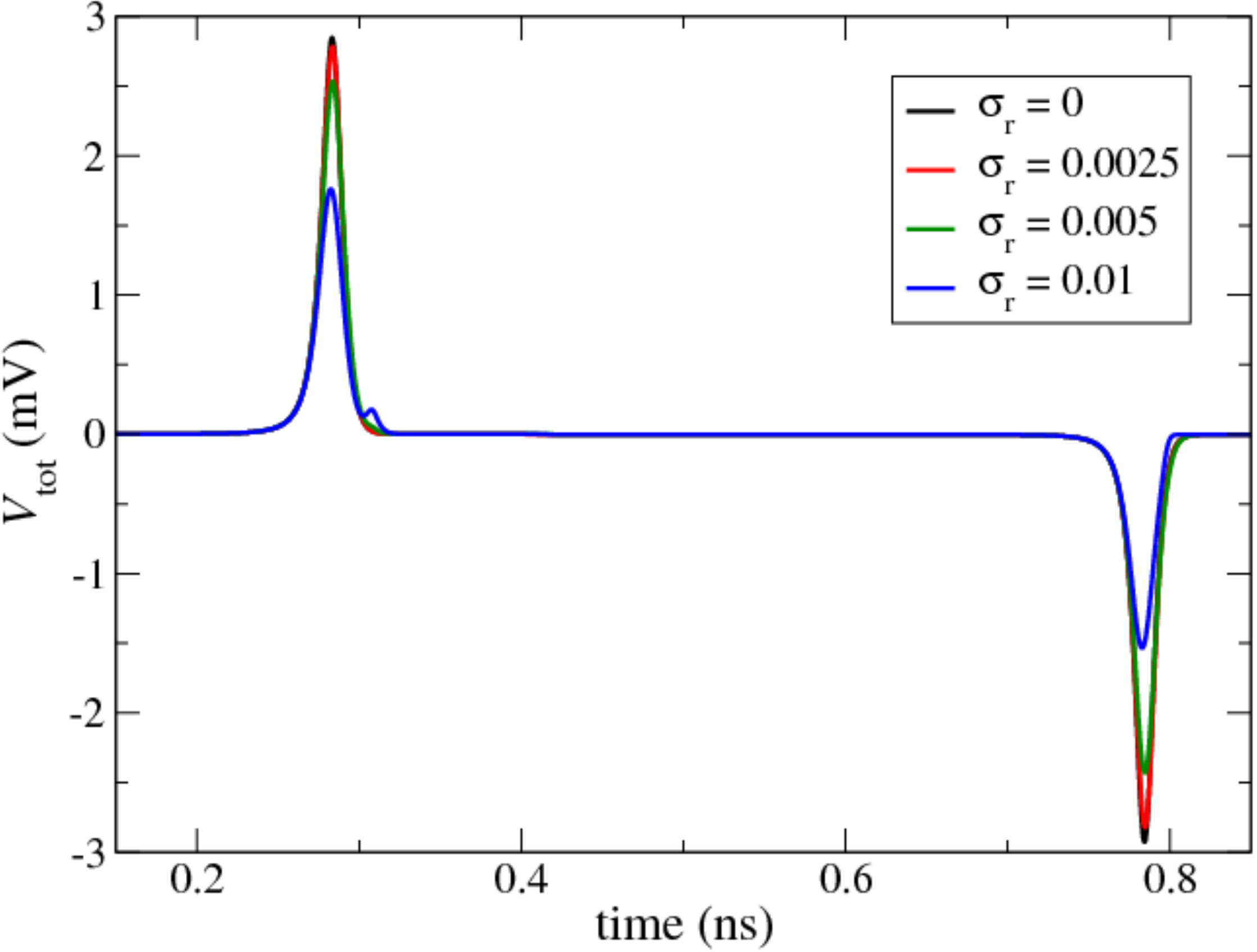}
\caption{(Color online) Behavior of a sequence of two typical voltage pulses generated by an array of $N=50$ SQUIDs made of Nb/AlOx/Nb junctions. The SQUID chain is characterized by an asymmetry parameter $r$ distribution which is gaussian and centered around $r_0=0.01$ with a standard deviation $\sigma_r$ [see Eq. \eqref{eq:AsymmetryDistribution}]. The driving frequency is $\nu$=1 GHz,  whereas the other SQUIDs parameters are the same as in the previous figures.} 
\label{fig:SQUID_Nb_errAsymmetries_effect_nu1}
\end{figure}

To overcome this problem, we assume that the SQUIDs are fabricated with a small \emph{preferential} asymmetry $r_0$, for instance $I_{c,2}<I_{c,1}$, which correspond to $r_0>0$.
We introduce a gaussian statistical distribution for the parameter $r$:
\be
r=r_0+\zeta_r\quad \text{with} \quad \mathcal{P}(\zeta_r)=\frac{1}{\sqrt{2\pi}\sigma_r}\,\text{exp}\left(-\frac{\zeta_r^2}{2\sigma_r^2}\right),
\label{eq:AsymmetryDistribution}
\ee
where $\zeta_r$ is a dimensionless parameter which quantifies the uncertainty on the SQUIDs asymmetry, being normally distributed around zero with variance $\sigma_r^2$, whereas $r_0=0.01$ is the chosen reference value for the SQUIDs asymmetry.
We have solved numerically the RCSJ dynamics of the linear array of SQUIDs following the same procedure outlined in the previous section. 
In Fig.~\ref{fig:SQUID_Nb_errAsymmetries_effect_nu1} we show how the \emph{typical}\footnote{As in Sec. \ref{sec:results_areas}, the typical voltage is the result of a statistical average over many realizations.} voltage pulses generated by an array of $N=50$ SQUIDs are altered by assuming different uncertainties $\sigma_r$ on the parameter $r$, up to a standard deviation of one percent. Notice that the main qualitative difference with respect to the previous cases, in which symmetric SQUIDs were considered, is that here the sequence of voltage pulses exhibits alternating signs. This feature was observed and explained in Ref.~\onlinecite{Solinas2014}: Its major consequence is that in the power spectrum the odd harmonics are predominant over the even ones.

\begin{figure}[t]
\centering
\includegraphics[width=0.9\columnwidth,  keepaspectratio]{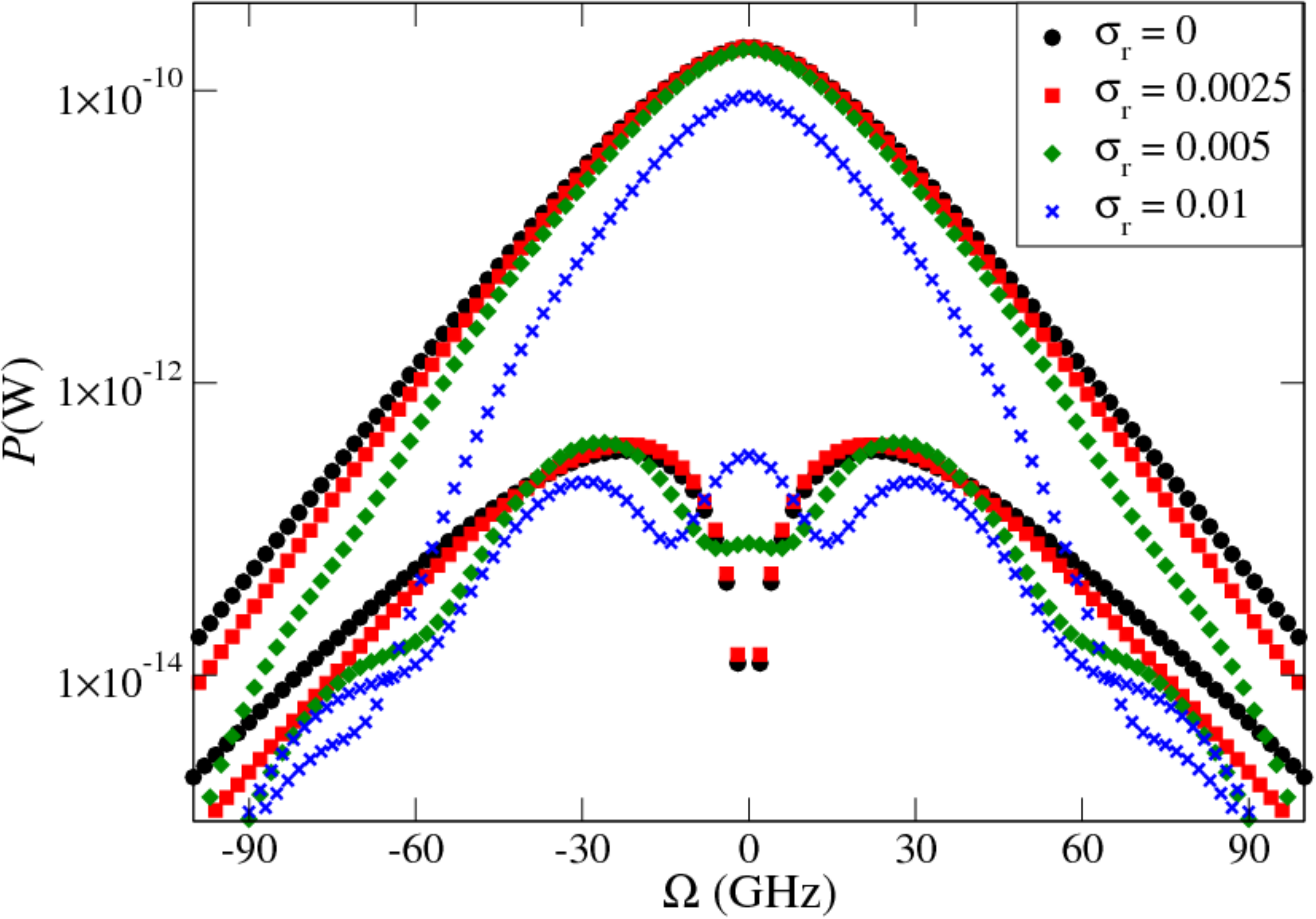}
\caption{(Color online) Power spectrum of the SQUID-Josephson radiation comb generator over a 50 Ohm transmission line, for different values of the standard deviation $\sigma_r$ of the asymmetry parameter distributions (centered around $r_0=0.01$). The calculation is performed for a $N = 50$ chain of Nb/AlOx/Nb SQUIDs subject to a $\nu$=1 GHz driving.} 
\label{fig:SQUID_Nb_errAsymmetries_power_nu1}
\end{figure}

Figure~\ref{fig:SQUID_Nb_errAsymmetries_power_nu1} shows the average power spectrum of the emitted radiation for an array of $N=50$ SQUIDs. 
We notice that increasing the uncertainty $\sigma_r$ on the asymmetry parameter reduces the power, especially at high frequency. 
Similarly to what we observed in Fig.~\ref{fig:SQUID_Nb_errAreas_power_nu1}, the non-dominant harmonics (in this case the even ones) show complex structure when increasing $\sigma_r$.
For $\sigma_r=0.01$, at high frequency ($\Omega\gtrsim 60$ GHz), the power associated to even harmonics becomes comparable or even larger than that associated to the odd ones.
\section{Experimental Feasibility}
\label{sec:feasibility}
In this final section we discuss the experimental feasibility of the setup, and we estimate its realistic performance when \emph{all} the parasitic effects studied so far are taken into account at once.
Some of the effects we are going to discuss were studied in Ref.~\onlinecite{Solinas2014}, so here we just review them briefly.

First of all, in our analysis so far we have neglected the coupling between the
SQUIDs via mutual inductance and/or cross capacitance and inductance of the superconducting wire. 
This condition, which basically relies only on the current conservation through each SQUID in the chain~\cite{Solinas2015}, implies that the dynamics of each SQUID is independent from the rest of the array, and it can be realized in practice by a suitable design choice.
As a consequence, the voltage at the extremes of the array scales as the number $N$ of SQUIDs. 
Accordingly, the \emph{intrinsic} power generated by the device (that is, the power delivered to an ideally infinite load) scales as $N^2$.
On the other hand, the \emph{extrinsic} power depends on the detection system used. In our case the JRCG array is supposed to be attached to a finite load, which effectively couples the dynamics of the SQUIDs: For realistic devices the extrinsic power is then found to scale as $N$, rather than $N^2$ (see related discussion in Sec.~\ref{sec:model}).
As shown in Sec.~\ref{sec:results}, this $N$ scaling is not a limitation in the region of tens of GHz, where sizable output power can be generated. Conversely at higher frequency, e.g. sub-millimeter region, the output power drops and the device design must be modified to compensate for this decrease. One possibility is to operate with more JRCG arrays arranged in a parallel configuration: In this case the contribution $P$ of each JRCG array would add up and the total power would be given by $P_\text{tot}=N_\text{par} P$, where $N_\text{par}$ is the number of SQUID arrays in parallel.

Another important issue concerns the way the emitted radiation propagates across the device.
When discussing the scaling of the power with the number $N$ of SQUIDs in the chain, we have implicitly assumed such radiation to propagate \emph{instantly} across the device.
Strictly speaking, this lumped-element model is justified if the propagation time $\tau_p$ of the radiation through the whole array is much shorter than the typical voltage pulse width, i.e. the voltage transient.  
This condition strongly depends on the specific values of the parameters, which in turn are set by the device fabrication, its design, and the materials used. All these can be optimized with the aim of decreasing the propagation time $\tau_p$.
In any case, this lumped model approximation does by no means set any sharp boundary condition on the working operation of the device. Even if $\tau_p$ was not \textit{much} shorter, but rather comparable with the voltage transient, the only consequence would be that the interference effects shall be taken into account. 
But since the device generates \textit{all} the harmonics of the fundamental frequency, some of them will be partially attenuated because of destructive interference, some others will be (almost) unchanged because of constructive interference.
Therefore the output signal may be attenuated at some specific frequencies, remaining unchanged at the others, but the device would still work and can be used if the specific output frequency we want to extract has enough output power.

\begin{figure}[b]
\centering
\includegraphics[width=0.9\columnwidth,  keepaspectratio]{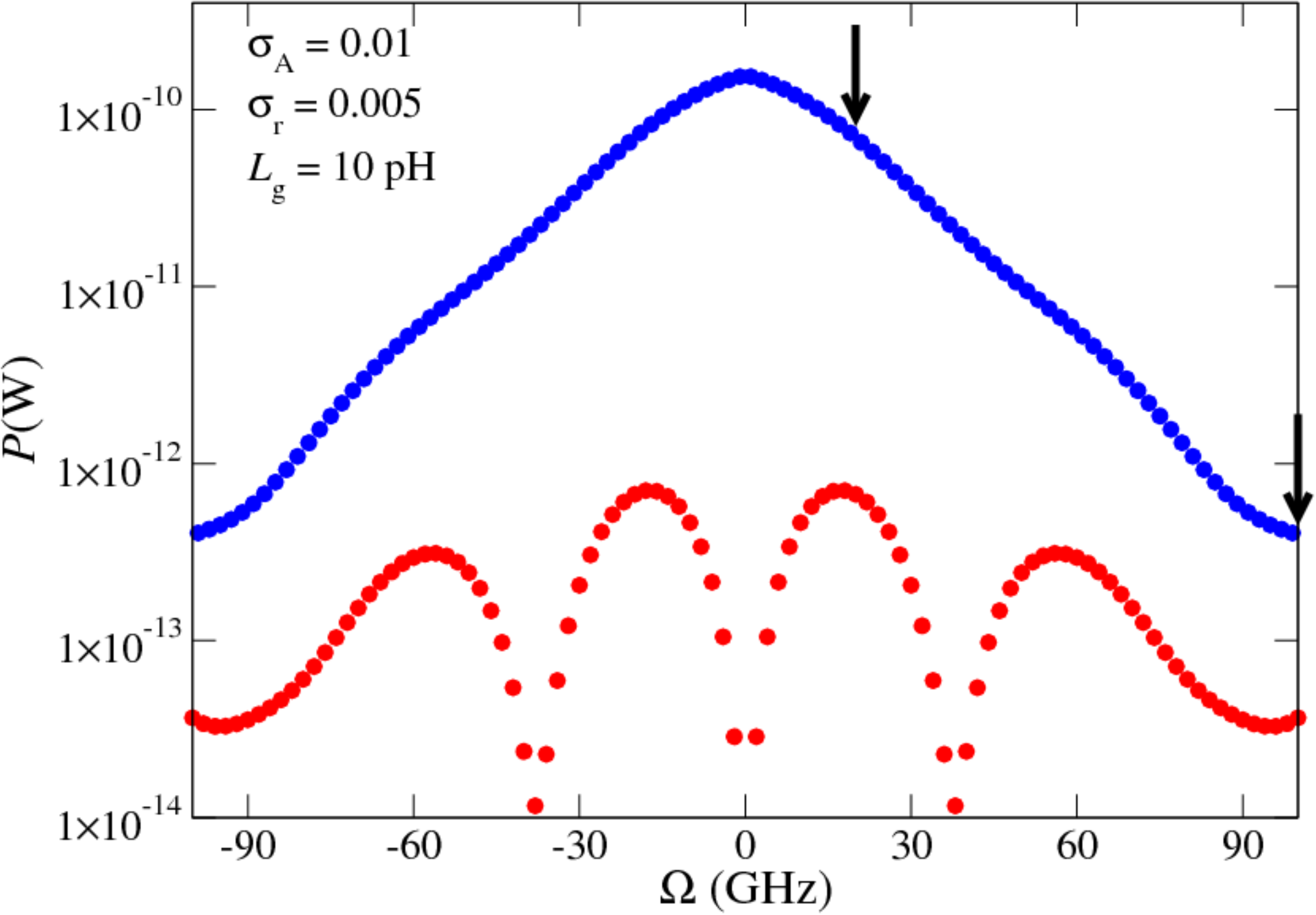}
\caption{(Color online) Power spectrum of a single realistic sample of a SQUID-based radiation comb generator over a 50 Ohm transmission line. We have assumed a geometrical inductance $L_g$=10 pH, and standard deviations $\sigma_A$=0.01 and $\sigma_r$=0.005 on the areas and the asymmetry parameter, respectively (the latter centered around $r_0=0.01$). The calculation is performed for a $N = 50$ chain of Nb/AlOx/Nb SQUIDs subject to a $\nu$=1 GHz driving. Blue and red symbols represent the odd and even harmonics, respectively, whereas the black arrows emphasize the frequency ranges around $20$ GHz and $100$ GHz.} 
\label{fig:SQUID_Nb_realistic_power_nu1}
\end{figure}

In Fig.~\ref{fig:SQUID_Nb_realistic_power_nu1} we show the estimated power spectrum generated by a single \emph{realistic} array of $N=50$ SQUIDs made of Nb/AlOx/Nb Josephson junctions. By ``realistic'' we mean subject to the fabrication errors discussed in Secs.~\ref{sec:results_areas} and \ref{sec:results_asymmetries}, having random (normally distributed) areas \emph{and} asymmetry parameters. Furthermore, we assume them to have a finite loop geometrical inductance $L_g\simeq 10$ pH (see Sec.~\ref{sec:results_inductance}). On the other hand, we do not consider any corrections due to their finite junction capacitance since we showed in Sec.~\ref{sec:results_capacitance} that they were completely negligible.
From the figure, we notice that this device is still able to provide an output power of about 0.1 nW around 20 GHz (corresponding to the 20-th harmonics of the driving frequency, see the corresponding black arrow). 
If we compare this to the results of Fig.~\ref{fig:SQUID_Nb_inductance_power_nu1}, we notice that the power in this frequency range is only slightly reduced, as a consequence of the errors on the areas and the asymmetry parameters.
A larger deterioration of the performance - of about two orders of magnitude - is otherwise expected at higher frequency (around $100$ GHz, see the corresponding black arrow).
Nevertheless, the device is still able to produce an output power between $0.1$ and $1$ pW in this range, which can be relevant for several applications.
All these considerations enforce the message that if the SQUIDs of the array can be fabricated with an accuracy of the order of $1\%$ on the areas and of $0.5\%$ on the asymmetry between the junctions, the expected performance is not altered significantly with respect to the ideal situation for frequencies around $20$ GHz.

Finally, we stress that all our analysis has been carried out at zero temperature, being more focused on the fabrication parasitic effects. The effects of thermal noise were indeed already addressed in Ref.~\onlinecite{Solinas2014} for a similar setup made of yttrium barium copper oxide (YBCO) Josephson junctions. In that case it was shown that its contribution was basically negligible, the signal to noise ratio being of the order of $10^3$ at a temperature of $4.2$ K. Hence, we do not expect a finite temperature to alter significantly the results presented in this paper.
\section{Conclusions}
\label{sec:conclusions}
In summary, we have discussed extensively several parasitic effects on the working operation of the SQUID-based radiation comb generator originally proposed in Ref.~\onlinecite{Solinas2014}.
Under certain conditions, we found that taking into account the finite loop geometrical inductance of the SQUIDs has a beneficial impact on the device performance, whereas the fabrication errors (uncertainties in the SQUIDs areas and asymmetries) tend to decrease it.
Also, in the range of parameters considered, we showed that a finite junction capacitance does not alter the results, meaning that the device operates always in the overdamped regime.\\
When all these effects are taken into account at once, we have estimated that a realistic array of $N=50$ SQUIDs made of Nb/AlOx/Nb junctions is able to deliver a power of $\sim 0.1$ nW around 20 GHz, and of $\sim 0.1-1$ pW around 100 GHz, to a standard load resistance of 50 Ohm. This may opens interesting perspectives in the realm of quantum information technology.\\
The device has room for optimization by modeling the geometry of the single junctions, the fabrication materials, the driving signal and the array design. For instance, besides SQUIDs made of tunneling junction considered in this work, one may investigate devices made of weak-link superconductor-normal metal-superconductor SNS junctions, such as Nb/HfTi/Nb Josephson junctions~\cite{Niemeyer2002,Koelle2013}.\\
Finally, the discussed implementation would have the advantage to be built on-chip  and  integrated  in  low-temperature  superconducting microwave electronics~\cite{Blais2004,Wallraff2004,Koch2007}.

\section*{Acknowledgments}
Stimulating discussions with C. Altimiras are gratefully acknowledged.
The work of  R.B. has been supported by MIUR-FIRB2013 -- Project Coca (Grant
No.~RBFR1379UX).
P.S. has received funding from the European Union FP7/2007-2013 under REA
Grant agreement No. 630925 -- COHEAT and from MIUR-FIRB2013 -- Project Coca (Grant
No.~RBFR1379UX). 
F.G. acknowledges the European Research Council under the European Union's Seventh Framework Program (FP7/2007-2013)/ERC Grant agreement No. 615187-COMANCHE for partial financial support.

\appendix
\section{Voltage spectrum and power}
\label{sec:app_power}
To test the performance of this radiation generator, we have calculated the power spectrum $P$ vs frequency $\Omega$. To this goal, first we have computed the Fourier transform of the voltage
\be
V(\Omega)=\int_0^{T}dt\,e^{i\Omega t} V(t).
\ee
The power spectral density (PSD) is then $PSD(\Omega) = 1/T|V(\Omega)|^2$. 
Finally, the power $P$ is calculated by integrating the PSD around the resonances $k\nu$ (where $\nu$ is the monochromatic driving frequency) and dividing for a standard load resistance of 50 Ohm. This is the power we would measure at a given resonance frequency with a bandwidth exceeding the linewidth of the resonance.

\section{Solution of the second-order RCSJ equation}
\label{sec:app_capacitance}

To study the dynamics of the SQUID phase $\varphi$ in Sec.~\ref{sec:results_capacitance}, we have used a downwind finite difference approach to discretize the derivatives in Eq.~\eqref{eq:RCSJ_dim3}, and the resulting equation implemented numerically is (in the dimensionless time notation):
\begin{align}
c&\,\frac{\varphi^{(i+1)}-2\varphi^{(i)}+\varphi^{(i-1)}}{d\tau^2}+\frac{\varphi^{(i)}-\varphi^{(i-1)}}{d\tau}+\cr
&+\alpha\left[\cos\phi^{(i)}\sin\varphi^{(i)}+r\,\sin\phi^{(i)}\cos\varphi^{(i)}-\delta\right]=0,
\end{align}
where $\varphi^{(i)}$ is the phase at time $\tau_i$, $\phi^{(i)}\equiv \pi\Phi(\tau_i)/\Phi_0$ is the reduced flux, 
$r$ is the asymmetry parameter of the SQUID, $c=2\pi\nu\,R_{\text{eff}}C$ is the reduced junction capacitance, $\alpha= I_+ R_\text{eff}/(\Phi_0 \nu)$ and $\delta=I_B/I_+$ is the dimensionless bias current.

\section{Statistical approach}
\label{sec:app_statisticalapproach}

In order to estimate the effects of imperfections in the SQUIDs fabrication, we have followed a statistical approach.
We describe here the procedure adopted in Sec.~\ref{sec:results_areas}, the one in Sec.~\ref{sec:results_asymmetries} being equivalent.

Given a certain value of the standard deviation $\sigma_A$, we have sampled an interval of width $8\sigma_A$ by introducing a number of bins $N_\text{bins}$.
We have then solved the RCSJ dynamics~\eqref{eq:RCSJ_dim3} for $N_\text{bins}$ values of $\zeta_{A,i}$ [corresponding to $N_\text{bins}$ values of areas $A_i$, according to Eq.~\eqref{eq:AreasDistribution}] taken as the centers of each bin. The computed voltage versus time $V_i(t)$ has been stored aside.

At this stage, to simulate the dynamics of an array, we have generated $N$=50 values of $\zeta_A$ taken from a random Gaussian probability distribution with zero mean and standard deviation $\sigma_A$, and to each one of these we have associated the voltage $V_i(t)$ corresponding to the closest value of $\zeta_{A,i}$, calculated and stored previously.

For an array of $N$ SQUIDs, under the hypothesis of independent SQUID dynamics (see Sec.~\ref{sec:feasibility}) the total voltage is simply the sum of all the voltages generated by each SQUID:
\be
V_{\text{tot}}(t)=\sum_{i=1}^N V_i(t).
\ee
Indeed the presence of the load, and the fact that it effectively couples the dynamics of the SQUIDs, has been taken into account by substituting the shunt resistance $R$ with $R_\text{eff}$ in the RCSJ equation, as discussed in Secs.~\ref{sec:model}.

Finally, this procedure has been iterated for a relatively large ($N_\text{real}$=10000) number of realizations of different arrays, and the typical voltage of an array has been defined as:
\be
V_\text{typ}(t)=\frac{1}{N_\text{real}}\,\sum_j V^{(j)}(t),
\ee
where the index $j=1\ldots N_\text{real}$ labels the $j$-th realization of an array.
We have done this, instead of simulating the dynamics of \emph{all} the SQUIDs of \emph{each} array many times, in order to reduce the computational burden, otherwise enormous.


\begin{thebibliography}{19}
\expandafter\ifx\csname natexlab\endcsname\relax\def\natexlab#1{#1}\fi
\expandafter\ifx\csname bibnamefont\endcsname\relax
  \def\bibnamefont#1{#1}\fi
\expandafter\ifx\csname bibfnamefont\endcsname\relax
  \def\bibfnamefont#1{#1}\fi
\expandafter\ifx\csname citenamefont\endcsname\relax
  \def\citenamefont#1{#1}\fi
\expandafter\ifx\csname url\endcsname\relax
  \def\url#1{\texttt{#1}}\fi
\expandafter\ifx\csname urlprefix\endcsname\relax\def\urlprefix{URL }\fi
\providecommand{\bibinfo}[2]{#2}
\providecommand{\eprint}[2][]{\url{#2}}

\bibitem{udem2002optical} T. Udem, R. Holzwarth and T. W. H{\"a}nsch, Nature \textbf{416}, 233 (2002).
\bibitem{cundiff2003} S. T. Cundiff and J. Ye, Rev. Mod. Phys. \textbf{75}, 325 (2003).
\bibitem{delhaye2007} P. Del'Haye, A. Schliesser, O. Arcizet, T. Wilken, R. Holzwarth and T. J. Kippenberg, Nature {\bf 450}, 1214 (2007).
\bibitem{hansch1999laser} T. H{\"a}nsch and H. Walther, Rev. Mod. Phys. \textbf{71}, S242 (1999).

\bibitem{bloembergen1977nonlinear} N. Bloembergen, \emph{Nonlinear spectroscopy}, vol.~\bibinfo{volume}{64}  (\bibinfo{publisher}{North Holland}, \bibinfo{year}{1977}).

\bibitem{hansch1994frontiers}
\bibinfo{author}{\bibfnamefont{T.~W.} \bibnamefont{H{\"a}nsch}}
  \bibnamefont{and} \bibinfo{author}{\bibfnamefont{M.}~\bibnamefont{Inguscio}},
  \emph{\bibinfo{title}{Frontiers in Laser Spectroscopy: Varenna on Lake Como,
  Villa Monastero, 23 June-3 July 1992}}, vol. \bibinfo{volume}{120}
  (\bibinfo{publisher}{North Holland}, \bibinfo{year}{1994}).

\bibitem{foreman2007remote}
\bibinfo{author}{\bibfnamefont{S.~M.} \bibnamefont{Foreman}},
  \bibinfo{author}{\bibfnamefont{K.~W.} \bibnamefont{Holman}},
  \bibinfo{author}{\bibfnamefont{D.~D.} \bibnamefont{Hudson}},
  \bibinfo{author}{\bibfnamefont{D.~J.} \bibnamefont{Jones}}, \bibnamefont{and}
  \bibinfo{author}{\bibfnamefont{J.}~\bibnamefont{Ye}},
  \bibinfo{journal}{Rev. Sci. Instrum.}
  \textbf{\bibinfo{volume}{78}}, \bibinfo{pages}{021101}
  (\bibinfo{year}{2007}).
	
\bibitem{DeWaele1969} A. T. A. M. De Waele, and R. De Bruyn Ouboter, Physica \textbf{41}, 225 (1969).
\bibitem{Barone1982} A. Barone, and G. Patern\`o, {\it Physics and Applications of the Josephson Effect} (John Wiley \& Sons, New York, 1982).
\bibitem{Solinas2014} P. Solinas, S. Gasparinetti, D. Golubev, and F. Giazotto, Scientific Reports \textbf{5}, 12260 (2015).
\bibitem{Solinas2015} P. Solinas, R. Bosisio, and F. Giazotto, J. Appl. Phys. \textbf{118}, 113901 (2015).
\bibitem{Tinkham2012introduction} M. Tinkham, {\it Introduction to superconductivity} (Courier Dover Publications, 2012).

\bibitem{giazotto2013coherent} F. Giazotto, M. J. Mart{\'\i}nez-P{\'e}rez, and P. Solinas, Phys. Rev. B \textbf{88}, 094506 (2013).
\bibitem{martinez2012nature}  F. Giazotto, and M. J. Mart{\'\i}nez-P{\'e}rez, Nature \textbf{492}, 401 (2012).
\bibitem{martinez2014quantum}  M. J. Mart{\'\i}nez-P{\'e}rez, and F. Giazotto, Nat. Commun. \textbf{5}, 3579 (2014).
\bibitem{gross2005applied} R. Gross, and A. Marx, Lecture on "Applied Superconductivity", \url{http://www.wmi.badw.de/teaching/Lecturenotes/}.
\bibitem{patel1999self} V. Patel, and J. Lukens, IEEE Trans. Appl. Supercond. \textbf{9}, 3247 (1999).




\bibitem{Niemeyer2002} D. Hagedorn, R. Dolata, F.-Im. Buchholz, and J. Niemeyer, Physica C \textbf{372}, 7 (2002).
\bibitem{Koelle2013} R. W\"olbing, J. Nagel, T. Schwarz, O. Kieler, T. Weimann, J. Kohlmann, A. B. Zorin, M. Kemmler, R. Kleiner, and D. Koelle, Appl. Phys. Lett. \textbf{102}, 192601 (2013).
\bibitem{Blais2004} A. Blais, R.-S. Huang, A. Wallraff, S. M. Girvin, and R. J. Schoelkopf, Phys. Rev. A \textbf{69}, 062320 (2004).
\bibitem{Wallraff2004} A. Wallraff, Nature \textbf{431}, 162 (2004).
\bibitem{Koch2007} J. Koch, T. M. Yu, J. Gambetta, A. A. Houck, D. I. Schuster, J. Majer, A. Blais, M. H. Devoret, S. M. Girvin, and R. J. Schoelkopf, Phys. Rev. A \textbf{76}, 042319 (2007).




\end{thebibliography}

\end{document}